\documentstyle[12pt]{article}

\def\be {\begin{eqnarray}}
\def\ee {\end{eqnarray}}

\def\vperp {v_\perp }
\def\vk {V_{\bf K}}

\begin{document}
\setcounter{page}{1}
\hfill {NORDITA-94/16 A/N/S}
\vskip 0.4in
\begin{center}
{\large\bf
Neutrino Emission from Dense Matter, \\
and Neutron Star Thermal Evolution\footnote{\it
To be published in the Proceedings of the NATO Advanced Study
Insitute on ``Lives of the
Neutron Stars", ed. A. Alpar, {\"U}. Kizilo{\u{g}}lu and J. van Paradijs
( Kluwer, Dordrecht, 1994 ). } }
\end{center}
\vskip 0.6in
\centerline{C. J. Pethick $^{1,2}$ and Vesteinn Thorsson$^{1}$ \\}
\vskip 0.4in
\centerline{\it $^{1}$ NORDITA, Blegdamsvej 17, DK-2100 Copenhagen \O, Denmark
}
\centerline{\it $^{2}$ Department of Physics, University of Illinois at
Urbana-Champaign,}
\centerline{\it 1110 West Green St., Urbana, Illinois 61801-3080, USA}
\vskip .6in

\centerline{\bf Abstract}
\vskip .5cm
\noindent
A brief review is given of neutrino emission processes in dense matter, with
particular emphasis on recent developments.  These include direct
Urca processes for nucleons and hyperons, which can give rise to rapid energy
loss from the stellar core without exotic matter, and the effect of band
structure on neutrino bremsstrahlung from electrons in the crust, which
results in much lower energy losses by this process than had previously been
estimated.
\par\vfill

\newpage

\section{Introduction}

The problem of the thermal evolution of neutron stars is a complex one, and in
investigating it theoretically one can identify a number of rather distinct
aspects.  The first set of questions concerns the problem of how thermal
energy is lost from the star.  One way in which this occurs is by emission,
from matter in the stellar interior, of neutrinos, and possibly of other
particles that can readily escape from the neutron star.  These aspects of the
problem are the ones we shall focus on in this article.  Another way is by
emission of thermal energy from the surface of the star.  The energy radiated
is transported to the surface by thermal conduction and radiative transport,
the former process being the most important one at high densities, and the
latter being the most important at low densities.  Heat flow to the stellar
surface is controlled by a bottleneck at relatively low densities, where
neither thermal conduction nor radiative transport are particularly effective.
These problems have been considered in detail by a number of authors, both in
the absence of magnetic fields{\cite{bottle1}}, and with the effects of the
magnetic field included{\cite{bottle2}}.

A second set of questions concerns mechanisms for heating the neutron star.
Among these one may name friction between superfluid and normal components of
the quantum fluids in the interior, if some of these are superfluid, and ohmic
losses due to the electrical currents in the star that sustain the star's
magnetic field.  Some of these processes will be considered in other
contributions to this volume.

A third set of questions concerns the spectrum of radiation emitted from the
stellar surface, which will be considered elsewhere in these
proceedings{\cite{radiation}}.  This is of great importance for understanding
observations of radiation from neutron stars{\cite{ogel}}, but fortunately the
total energy radiated from the stellar surface is largely independent of
processes in the last few photon mean free paths in the atmosphere, and
therefore one can calculate the total energy radiated without knowing the
physics that determines the details of the stellar spectrum, just as is the
case in ordinary stars.

Ever since neutron stars became objects of serious study by physicists over a
quarter of a century ago it has been realized that the observation of thermal
radiation from neutron stars provides a way of probing the interior
temperature of the star, and therefore has potential for giving valuable
information about the nature of matter in the stellar interior.  Theoretical
arguments alone are insufficient to establish exactly what state of matter
exists in the cores of neutron stars, and therefore information that can be
gleaned from astronomical observations could play an important role.  As we
have heard in other talks at this meeting{\cite{ogel,xray}} techniques for
making observations in the X-ray and soft ultraviolet ranges of the spectrum
have now been developed to such a degree that this dream is close to being
realized.

We begin by summarizing the theoretical situation as of 1990, when the
previous workshop in this series on neutron stars and related phenomena was
held in Agia Pelagia.  If a neutron star were observed to cool slowly, on a
timescale of order $1 yr/T_9^4$, where $T_9$ is the core temperature in units
of $10^9$ K, this would be a clear sign of the core being composed of normal
matter (a mixture of neutrons, protons, and electrons, with possible
admixtures of other constituents), while if the neutron star were observed to
cool rapidly, on a timescale of order $1 min/T_9^4$, this would indicate that
matter in the core was in an exotic state, such as quark matter, or a Bose
condensate of pions or kaons.  The reason for the very different timescales is
that it was argued that in ordinary matter, the dominant neutrino emission
process is the so-called modified Urca process, first discussed by Chiu and
Salpeter{\cite{chiu}}, in which the two reactions

\be
n + n \rightarrow n +p +
e^- + \bar\nu_e \,\,\, {\rm and} \,\,\, n +p + e^- \rightarrow n + n +\nu_e
\label{modurca}
\ee
occur in
equal numbers.  These reactions are just the usual processes of neutron beta
decay and electron capture on protons,

\be
n \rightarrow p + e^- + \bar\nu_e \,\,\, {\rm and} \,\,\,
p + e^- \rightarrow n +\nu_e \,\,\, ,
\label{dirurca}
\ee
with the addition of an extra bystander neutron.  They produce
neutrino-antineutrino pairs, but leave the composition of the matter constant
on average.  The rate at which modified Urca reactions occur per unit volume
varies as $T^7$, a result that can easily be understood from phase space
arguments.  At temperature $T$, the energy available to a neutrino is of order
$k_B T$, and therefore, since the neutrinos are nondegenerate and can freely
escape from the star, the phase space volume accessible is a sphere in
neutrino momentum space, of radius $\sim T$.  Consequently the phase space
factor is proportional to $T^3$.  The other particles participating in the
modified Urca reaction are degenerate fermions, and therefore the number of
states (either initial or final) is proportional to the volume of a shell of
states with a width in energy $k_B T$ about the Fermi surface, and therefore
each of these fermions gives a factor $T$ to the phase space factor.  However,
the energies of the particles are constrained by the condition of energy
conservation, and therefore the energies of only 4 of the 5 degenerate
fermions are
independent.  As a result, the degenerate fermions give a factor $T^4$ to the
phase space factor.  Combining this with the factor for the neutrino phase
space, we see that the rate of the modified Urca process should vary as $T^7$.
For neutron star thermal evolution, the relevant quantity is the rate of {\it
energy} emission, $\dot E$ which, since the neutrino energy is $\sim k_B T$,
varies as $T^8$.  The thermal energy, $E_{th}$ of degenerate fermions varies
as $T^2$ at low temperatures, and consequently the cooling time,
\be
\tau = \frac{E_{th}}{\dot E} \,\,\, ,
\label{tau}
\ee
which is determined locally, varies as $T^{-6}$, as given
above.

In the case of exotic states it is possible for processes of the type

\be
f_1  \rightarrow  f_2 + e^- +\bar\nu_e \,\,\, {\rm and} \,\,\,
f_2 + e^- \rightarrow    f_1 + \nu_e
\label{genurca}
\ee
to occur.  Here $f_1$ and $f_2$ are two fermions, whose character depends
on the specific exotic state considered. From arguments analogous to those
given above for the modified Urca process, one can see that the rate of
neutrino energy loss varies as $T^6$, and therefore the characteristic
cooling time varies as $T^{-4}$, in agreement with the result stated earlier.
For a discussion of the relevant reactions for specific exotic states, we
refer to Ref. \cite{rmp}.

Over the past few years it has become clear that there are a number of
possibilities for processes of the type (\ref{genurca}) to occur in dense
matter, even in the absence of exotic states.  These so-called ``direct Urca
processes'' will be discussed in Sec. 2.

The arguments we have made for the temperature dependence of reactions that
produce neutrinos depend on the spectrum of elementary excitations being
smooth in the vicinity of the Fermi surface.  Should matter become superfluid,
or superconducting, by pairing of particles, as in metallic superconductors,
gaps can open up at the Fermi surface, and these will suppress neutrino
emission at temperatures less than or of the order of the transition
temperature to the paired state.  If the effects of superfluidity in the core
of the star are sufficiently pronounced, neutrino emission from the crust of
the neutron star could be the dominant energy loss mechanism.  Recent work on
what was generally regarded as the dominant neutrino production mechanism in
the crust, bremsstrahlung of neutrino pairs by electrons, has shown that at
temperatures of order $10^{10}$K this process is much less effective than
previously estimated\cite{pt}.  The reason for this suppression is the band
structure of electron states resulting from the motion of the electrons in the
periodic lattice of nuclei, an effect we shall describe in the Sec. 3.

In Sec. 4 we shall address a number of unsolved problems.  These include the
reliability of estimates of the modified Urca process, and some processes that
can occur in neutron star crusts as a consequence of the aspherical nuclear
shapes described elsewhere in this volume\cite{pr}.  Concluding remarks are
made in Sec.5.

\section{Direct Urca Processes}

One of the simplest possible ways of generating neutrinos and antineutrinos in
dense matter is the pair of processes (\ref{dirurca}), and we begin by
investigating the their kinematics.  In degenerate matter, the condition of
energy conservation, together with the requirement that final states for
fermions must be unoccupied, lead to the conclusion that particles
participating in reactions must lie within an energy $\sim k_B T$ of their
respective Fermi surfaces.  The momenta of the degenerate fermions must
therefore be close to their Fermi momenta, and the neutrino momentum is of
order $k_B T/c$, which at low temperatures may be neglected compared with the
Fermi momenta.  The condition for momentum conservation therefore amounts to
the requirement that it be possible to construct a triangle from the three
Fermi momenta.  In the canonical view, neutron star matter at around nuclear
density consists mainly of neutrons with a small fraction of protons, together
with an equal number of electrons to ensure charge neutrality.  The density of
a particle of species $i$ is proportional to $p_i^3$, where $p_i$ is its Fermi
momentum, and therefore the Fermi momenta of protons and electrons are equal
if there are no other charged constituents.  If the proton fraction is of
order 1\%, as was estimated to be the case in the the 1960's, when microscopic
mechanisms for neutrino production in dense matter were first investigated,
one can see that $p_p/p_n=p_e/p_n \simeq 1/5$. Thus one can see that
it is impossible to satisfy the triangle inequality, and the direct Urca
process could not occur in dense matter.  On the basis of this argument, Chiu
and Salpeter were led to investigate other possibilities, in particular the
modified Urca process mentioned above, which acquired the status of the
``standard'' process for neutron star cooling.

\subsection{The Nucleon Process}

It is interesting to enquire just how large a proton fraction is necessary for
the direct Urca process to proceed.  If protons and electrons are the only
charged constituents, it is easy to see that, since the electron and proton
Fermi momenta are equal, the smallest proton Fermi momentum for which one can
construct a triangle from the Fermi momenta is one half of the Fermi momentum.
Thus the proton to neutron ratio is 1/8, and the ratio of protons to nucleons
is 1/9, or just over 11\%.  This calculation was made as long ago as 1981 by
Boguta{\cite{bog}} in a paper that went largely unnoticed by the neutron-star
community, and the same arguments were made a decade later in
Ref.{\cite{lpph}}, where the rate of the process was also calculated for
proton fractions in excess of the threshold value.  If one allows for muons as
well as electrons, the threshold proton fraction is slightly higher:  in
particular, when the electron chemical potential is much larger than the muon
rest mass energy, it is $\simeq 0.148$.

While such proton concentrations were regarded as unrealistically high in the
mid 1960's, the situation is less clear today.  Using methods that describe
the nucleon-nucleon interaction by a potential, Wiringa, Fiks, and
Fabrocini{\cite{wff}} find proton fractions that lie below the threshold
value, although for some interactions the proton fraction is rather close to
the threshold value.  It is important to note that proton fractions are not
well determined theoretically, because they depend on, among other things, the
isospin dependence of the three-body interaction, which is not well
characterized.  Models of dense matter based on mean field theory (see, e.g.,
Ref.\cite{chin}) tend to give larger proton fractions than do the potential
models, and
the proton fraction in many cases exceeds the threshold value.  To determine
whether or not the direct Urca process for nucleons is a realistic possibility
in dense matter, more reliable estimates of proton fractions are required.

Should the kinematical constraints for the process (\ref{dirurca}) be
satisfied, the cooling time (\ref{tau}), will be given by

\be
\frac{1}{\tau_{\rm Urca}} =
\frac{457}{3360 \pi}
\frac{G_F^2 \cos^2{\theta_C} (1+3g_A^2) }{ \hbar^7 c^6}
\frac{m_n c}{p_n}
\mu_e (k_B T)^4     \,\,\, ,
\label{timescale}
\ee
where the thermal energy, $E_{th}$, has been approximated by that for neutrons
alone\cite{rmp}.
Here, $G_F$ is the weak interaction coupling constant, $\theta_C$ the
Cabibbo angle, $g_A$ the axial vector coupling constant, and $m_n$ the
neutron rest mass.

One interesting possibility that has been explored recently is that the
appearance of a Bose condensate of kaons would lead to an increase of the
proton fraction, which, if it were sufficiently large, could result in the
kinematic condition for the direct Urca process for nucleons being
satisfied{\cite{kaons}}.  If a $K^-$ condensate appears, it is less
costly energetically to add negative charge than it is in the absence of the
condensate.  Since bulk matter is electrically neutral, this implies that in
equilibrium there will be more positive charge, i.e. protons, than there
would be in the absence of the condensate.  A competing effect is that when a
condensate appears, the electron fraction tends to decrease, because some of
the negative charge resides in the condensate, but for most models of dense
matter that have been investigated to date, this effect is more than
outweighed by the increase in the proton fraction.
Should the proton concentration exceed the threshold value for the direct
Urca process, the cooling time in the presence of a kaon condensate would
be $\tau_{KUrca}=\sec^2(\theta/2) \tau_{Urca}$,
where $\theta$ is angle that measures the amplitude of the condensate.
Even if the proton concentration is below the direct Urca threshold,
alternative processes are possible, such as (\ref{genurca})
with $f_1$ and $f_2$ both representing {\it neutrons}, modified due to
the presence of a condensate.
However, since such processes do not conserve strangeness,
the corresponding
cooling time, $\tau=4\cot^2\theta_C\csc^2\theta\tau_{Urca}$, is typically much
greater than $\tau_{KUrca}$.

\subsection{Processes for Hyperons and Isobars}

Dense matter may well contain particles other than the ones that we have
discussed up to now.  These include hyperons and resonances, which, while they
decay in the lab, can exist stably in dense matter because possible
states for the decay products are blocked by the Pauli principle.
As was pointed out in Ref.\cite {pplp}, such particles could also participate
in direct Urca processes.  Possible particles include $\Sigma^-$, and
$\Lambda$ hyperons, and
$\Delta$ isobars, and possible pairs of Urca processes include
\be
\Sigma^-  \rightarrow  n + e^- +\bar\nu_e \,\,\, {\rm and} \,\,\,
n + e^- \rightarrow    \Sigma^- + \nu_e  \,\,\, ,
\label{sigma}
\ee
\be
\Lambda \rightarrow p + e^- + \bar\nu_e \,\,\, {\rm and} \,\,\,
p + e^- \rightarrow \Lambda +\nu_e \,\,\ ,
\label{lambda}
\ee
\be
\Sigma^-  \rightarrow  \Lambda + e^- +\bar\nu_e \,\,\, {\rm and} \,\,\,
\Lambda+ e^- \rightarrow    \Sigma^- + \nu_e \,\,\, ,
\label{siglam}
\ee
and
\be
\Delta^-  \rightarrow  n + e^- +\bar\nu_e \,\,\, {\rm and} \,\,\,
n + e^- \rightarrow    \Delta^- + \nu_e \,\,\, .
\label{delta}
\ee

For these processes to occur, the Fermi momenta of the participating particles
must satisfy the triangle inequalities, just as in the case of the nucleon
process.  One interesting conclusion is that the process (\ref{lambda}) can
take
place for quite low concentrations of $\Lambda$ hyperons:  if matter consisted
only of neutrons, protons, electrons, and $\Lambda$ hyperons, the threshold
concentration of $\Lambda$ hyperons would be zero,
and for more realistic models of
dense matter, the threshold fraction of $\Lambda$ hyperons to nucleons is of
order one
part in a thousand.  Again, in order to evaluate whether any of these hyperon
and isobar processes are realistic possibilities, better estimates of the
composition of dense matter are needed.  Should the processes be allowed
kinematically, the characteristic cooling time is similar to the
nucleon direct Urca time, Eq.5, increased by a factor $\cot^2\theta_C$ when
the process involves a strangeness change, as reactions (6) and (7) do.

\section{Neutrino Pair Bremsstrahlung by Electrons}

The basic process is emission of a neutrino-antineutrino pair accompanying
scattering of an electron by ions in the crust of a neutron star.  The rate of
the process has previously been considered in many articles {\cite{brems}},
all of which share the common feature that the electron-nucleus interaction is
treated in first-order perturbation theory.  These estimates suggested that,
depending on what is assumed about the neutron star model, energy emission by
neutrino bremsstrahlung in the crust could be competitive with the
modified Urca process in the core if the core were made of normal matter,
while the crust bremsstrahlung process could dominate energy loss if the
nucleons in the core underwent a transition to a superfluid and/or
superconducting state.

The basic process is shown in Fig. 1a, where the double lines represent an
electron moving in the Coulomb potential of the nuclei, and Fig. 1b shows the
diagrams that contribute in the usual approximation in which the
electron-nucleus interaction is treated perturbatively.  The dependence of the
rate of energy emission on temperature can be found as we did earlier for the
modified Urca process.  Phase space gives a factor of $T^3$ for each of the
neutrino and antineutrino, a factor of $T$ for each of the incoming and
outgoing electrons, and a factor of $1/T$ to take into account the fact that,
because energy must be conserved in the process, one of the particle energies
is not an independent variable.  On the basis of phase space alone, one would
thus expect the energy emission rate, which contains an extra factor $T$ for
the energy of the neutrino and antineutrino, to vary as $T^8$, just as the
modified Urca process does.  However, in the contribution to the matrix
element corresponding to the perturbation theory diagrams in Fig. 1b there is
an intermediate energy denominator where the electron is off the energy shell
by an amount proportional to the momentum of the neutrino pair, $\sim T$.
Thus in the calculation of the rate, the square of the matrix element gives an
extra factor $T^{-2}$, and the rate of energy emission is proportional to
$T^6$.

{}From the above considerations one can easily see that the perturbation
expansion is really an expansion in terms of the strength of the electron-ion
interaction compared with the energy denominator $\sim T$.  The strength of
the Coulomb interaction with the static lattice is of order $|\vk|$, where
$\vk$ is the electron-ion matrix element, which depends on the reciprocal
lattice wavevector, $\bf K$, describing the scattering process.  The crystal
potential gives rise to electronic band structure with splittings between
bands whose magnitude is $2|\vk|$.  For point-like nuclei, the splitting for
the lowest reciprocal lattice vector is $\simeq 0.018(Z/60)^{2/3}\mu_e$, where
$\mu_e$ is the electron chemical potential, and $Z$ is the atomic number of
the nucleus.  Since $Z$ is estimated to be close to 60 and $\mu_e$ is $\sim$
75 MeV at the highest densities at which spherical nuclei exist in the crust,
band splittings can be as large as 1 MeV. This shows that perturbation theory
is inapplicable at temperatures below about $10^{10} {\rm K}$, since the
dimensionless parameter in the perturbation series is then larger than unity.

Recently we have calculated the rate of the bremsstrahlung process without
making the perturbation expansion\cite{pt}.  The rate of the process is
calculated from
the diagram shown in Fig. 1a, with the weak-interaction matrix elements being
ones for band states, where the electron-ion interaction is included to
arbitrarily high order.   An analytical result has been obtained in the low
temperature limit, in which $k_B T \ll |\vk|$:

\be
{\dot E}_<
 &=& \frac{2 G_F^2}{ 3 \pi^\frac92 \hbar^{10} c^9 }
\frac{ C_A^2 + C_V^2 }{2}
\mu_e
(k_BT)^\frac72
\sum_{\bf K}
\frac{(1-\vperp)^{\frac 12} }{
\vperp^{\frac 52}  ( 1 + \vperp)   }
|\vk|^\frac92
e^{-\frac{1}{k_BT} \frac{2|\vk|}{1+v\perp} }  \,\,\, .
\label{lowtemp}
\ee
Here,
$C_A = 1/2$, $C_V = 1/2 + 2\sin^2\theta_W$ in terms of the weak mixing angle
$\theta_W$,
and $v_{\perp} = \sqrt{1 - (K/(2k_F))^2}$,
with $k_F \simeq \mu_e/(\hbar c)$ being the Fermi wavenumber.
The total emission rate from all types of neutrinos is obtained
by replacing
$ C_V^2 + C_A^2$ by
$C_A^2 + C_V^2 + 2 ( (1-C_A)^2 + (1-C_V)^2 ) $.
This shows that the emission rate falls exponentially at temperatures below
the band splitting for the particular reciprocal lattice vector considered.

We have not yet performed detailed numerical integrations to evaluate the
neutrino emission rate when band splittings are comparable to the thermal
energy, but we have constructed a simple formula which interpolates between
the perturbation theory result, ${\dot E}_>$, which is valid at high
temperatures, and Eq.(\ref{lowtemp}), which holds at low temperatures.
Results for conditions
appropriate for the highest density at which nuclei are roughly spherical are
shown in Fig. 2, where we compare the energy emission rate according to the
interpolation formula with ${\dot E}_>$.
We see that at temperatures below about $10^{10} {\rm K}$, the
reduction in the neutrino emissivity is more than a factor 2, and below $10^9
{\rm K}$ it is more than an order of magnitude.

What these results demonstrate is that for perturbation theory calculations of
the rate of the bremsstrahlung process to be applicable it is necessary that
band splittings be less the thermal energy, and not just less than the Fermi
energy, as one might expect at first sight.  This circumstance is a
consequence of the small energy denominators that arise in the perturbation
expansion.

\section{Open Questions}

The rate of the modified Urca process has been estimated by a number of
authors, and the one that is generally quoted is that by Friman and
Maxwell\cite{fm}.  A perturbation theory diagram for a typical process is
shown in Fig. 3a.  Some years ago Voskresenski\u{\i} and Senatorov\cite{vs}
pointed out two effects that had not been taken into account in earlier
estimates.  The first is that the nucleon-nucleon interaction is modified by
the presence of the dense medium.  In calculations of the rate of the modified
Urca process, the long range part of the interaction, due to exchange of a
single pion, dominates the rate.  The pion field is modified by the medium,
which in turn changes the long-range part of the nucleon-nucleon interaction.
A second effect is that the exchanged pion, rather than one of the nucleons,
can undergo a weak interaction process, as shown in Fig. 3b.  The latter is
an example of what is usually referred to as an ``exchange current effect''.
The estimates of Ref.\cite{vs} suggest that these two
effects could lead to an enhancement of the rate of the modified Urca process
by as much as several orders of magnitude.  However, the results are very
dependent on the specific assumptions made, and more detailed estimates should
be made.

Another set of questions concerns neutrino emission in the crusts of neutron
stars.  In Sec.3 we considered bremsstrahlung of neutrino pairs by electrons
moving in the static Coulomb field of the nuclei, and showed that it is
suppressed exponentially at low temperatures.  This indicates that the
phonon-assisted process, in which lattice vibrations generate neutrino pairs,
will be the dominant mechanism for bremsstrahlung of neutrino pairs by
electrons, since it has a power-law temperature dependence.  This process
deserves to be reexamined.  One may also ask whether, in the crust, there are
neutrino generating processes in which nucleons can participate\footnote{Such
processes have previously been considered for the bulk interior, where
neutrino pair bremsstrahlung can be produced in nucleon-nucleon
collisions{\cite{sutherland,fm}}, but for normal matter it is estimated to be
less effective than the modified Urca process.}. One of the
qualitatively new features of the states with non-spherical nuclei that are
discussed in another contribution to this volume\cite{pr} is that both
neutrons and protons have continuous energy spectra in the vicinity of the
respective Fermi energies.  As a consequence, processes involving nucleons are
not suppressed exponentially at low temperatures as they are in finite nuclei,
where the spacing between different single-particles states is finite.
Leinson\cite{leinson} has pointed to the possibility of neutrino pairs being
produced by scattering of nucleons from inhomogeneities in the nuclear medium.
Leinson considered the case of matter at relatively high temperatures, and he
assumed that the nuclear charge distribution corresponded to a collection of
bubbles at random positions, and he concluded that the process could be a
significant source of energy loss during the very early life of a neutron
star.

\section{Conclusion}

As our discussion shows, there have been a number of advances in our
understanding of neutrino emission from dense matter over the past few
years.  New processes have been discovered, and some old ones have
been found to have rates significantly different from earlier estimates.

One new result is the discovery  of a number of possible processes in
dense matter that can give rise to rapid cooling without the need for an
exotic state.  Consequently one can no longer argue that rapid cooling is a
unique signature of an exotic state of matter in the interior of a neutron
star.  Whether or not these new processes take place depends crucially on the
composition of neutron star matter at supernuclear densities, and
therefore it is important in future to attempt to narrow down the range of
possible compositions of dense matter.

Bremsstrahlung of neutrino pairs in the crusts of neutron stars has been shown
to be less important than previously estimated for two reasons:  first, the
rate of the process is suppressed by band-structure effects at temperatures
below $\sim 10^{10}$ K, and second, the amount of matter in the crust of a
neutron star is now estimated to be considerably less than was given by
earlier calculations.  With respect to the crust, one topic for future
investigation is whether, in phases with nonspherical nuclei, the neutral
current process proposed by Leinson, or its charged current analog\cite{lrp},
are quantitatively important for neutron star cooling.

The work of Voskresenski\u{\i} and Senatorov provides motivation for a
thorough reinvestigation of the modified Urca process, with allowance for the
knowledge of pion physics that has been accumulated over the last two decades.
The modified Urca process is much more sensitive to strong interactions,
since, unlike the direct Urca processes, its rate depends explicitly on the
nucleon-nucleon interaction, and not just implicitly, through the composition.

Kaon condensation, were it to occur, would have important consequences for
neutron star cooling, and more generally for the composition of dense matter.
Calculations that take into account more physics than has been included
to date are needed in order to assess how realistic a possibility
kaon condensation is.  Among these effects are the possible effects of
resonance states, such as $\Lambda(1405)$, and higher-body contributions to
kaon-nucleon interactions.

One topic that we have not touched on in detail in the text is superfluidity
and superconductivity of nucleons.  Among recent developments are new
evaluations of gaps\cite{gaps}, and detailed calculations of the suppression
of the direct Urca rate\cite{yak} for both isotropic and anisotropic gaps.
Among other more speculative possibilities that we have not been able to cover
in this brief review are cooling by emission of axions, which allows one to
obtain bounds on axion properties{\cite{iwamoto}}, and the direct Urca process
occurring in phases consisting of coexisting quark matter and nuclear
matter\cite{hps}.

\vskip 0.5cm
This research was supported in part by grants from
the US National Science Foundation (NSF PHY91-00283), and from NASA
(NAGW-1583).

\newpage
\section*{Figure Captions}

\noindent
FIG.1. a) The basic bremsstrahlung process.
The double line is the
propagator for a band electron. b) The process in first-order perturbation
theory. The cross denotes an electron-lattice interaction, and the
propagators are free ones.

\medskip \noindent
FIG.2.
Energy emission rate according to the interpolation formula compared with the
high temperature limit,
$\dot{E}_>$,
as a function of temperature.

\medskip \noindent
FIG.3.
a) Perturbation-theory diagram for a typical contribution to the modified Urca
process.  The wavy line represents a weak interaction, while the dashed line
represents a strong interaction. b) Contribution to the modified Urca process
in which exchanged pion emits leptons.

\end{document}